# A Calibration Method for Indirect Time-of-Flight Cameras to Eliminate Internal Scattering Interference

Yansong Du[1], Jingtong Yao[1], Yuting Zhou[1], Feiyu Jiao[1], Zhaoxiang Jiang[2,3], and Xun Guan[1,4]

[1]Tsinghua Shenzhen International Graduate School, Tsinghua University, Shenzhen, 581055, China

[2]Guangdong Laboratory of Artificial Intelligence and Digital Economy (SZ), Shenzhen, 518060, China

[3]jiangzhaoxiang@gml.ac.cn

[4]xun.guan@sz.tsinghua.edu.cn

**Abstract** In-camera light scattering is a typical form of non-systematic interference in indirect Time-of-Flight (iToF) cameras, primarily caused by multiple reflections and optical path variations within the camera body. This effect can significantly reduce the accuracy of background depth measurements. To address this issue, this paper proposes a calibration-based model derived from real measurement data, introducing three physically interpretable calibration parameters: a normal-exposure amplitude influence coefficient, an overexposure amplitude influence coefficient, and a scattering phase shift coefficient. These parameters are used to describe the effects of foreground size, exposure conditions, and optical path differences on scattering interference. Experimental results show that the depth values calculated using the calibrated parameters can effectively compensate for scattering-induced errors, significantly improving background depth recovery in scenarios with complex foreground geometries and varying illumination conditions. This approach provides a practical, low-cost solution for iToF systems, requiring no complex hardware modifications, and can substantially enhance measurement accuracy and robustness across a wide range of real-world applications.

## 1. Introduction

Indirect Time-of-Flight (iToF) cameras, known for their real-time performance, compact design, easy integration, and cost-effectiveness, have become widely used in 3D sensing applications such as industrial calibration [1], ambient monitoring [2], and automated positioning [3]. Among available techniques, Amplitude Modulated Continuous Wave (AMCW) is the most common implementation [4]. However, despite notable advances, iToF depth accuracy remains limited by various sources of error, especially internal scattering interference [5]. When a foreground object is positioned close to the iToF camera, its strong reflected light may enter the camera and undergo multiple reflections between the lens and the sensor surface, causing the signal to spread across adjacent pixels instead of focusing correctly [6]. This scattering contaminates the background depth, particularly when the foreground signal dominates, leading to significant degradation in measurement accuracy, as illustrated in Figure 1.

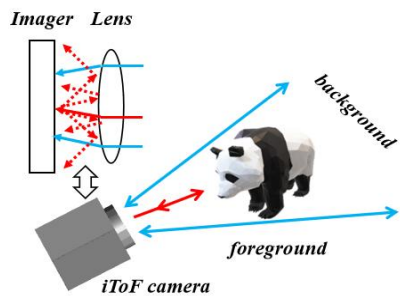

Fig. 1. Schematic Diagram of Internal Scattering Interference Imaging.

To address internal scattering interference in iToF cameras, a variety of methods have been proposed, such as lens and sensor surface coating to reduce reflections [7-8], device-specific error modeling [9], spatially variant point spread function (PSF) correction [10], and treating the scattering process as a convolution problem followed by compensation optimization [11]. Some studies also focus on hardware-level improvements, including redesigning CMOS sensor architectures [12] or adjusting modulation waveforms to suppress unwanted reflections [13]. In addition, feature alignment techniques like SIFT and ICP [14], as well as artifact prediction frameworks in multi-camera systems [15], have been explored. Many of these methods also rely on prior information such as target reflectivity [16] or imaging distance [17] to model scattering behavior. However, these factors can vary significantly in real-world applications, making unified modeling difficult and limiting adaptability. While deep learning has shown promising results in related ToF tasks, its application to internal scattering remains limited due to the lack of ground truth data, the complexity of modeling light propagation, and the poor generalization of models trained on synthetic datasets [18-23]. More importantly, most existing methods fail to explicitly model the optical coupling between internal scattering and the lens–sensor structure, which is the fundamental source of this issue.

This work presents a calibration method based on real measurement data and physically interpretable parameters to address the limitations of existing internal scattering correction techniques in iToF cameras. A compact parametric model is constructed to describe how foreground illumination interferes with background depth measurements. We fix the background distance and lighting environment, and acquire multiple sets of image data by varying the position of the foreground object (i.e., its distance from the lens) and the exposure time. These variables are carefully selected to ensure the presence of internal scattering effects while keeping them within the correctable range of our proposed method. Based on these observations, we extract the scattering-affected regions and estimate three calibration parameters: the normal-exposure amplitude influence parameter (reflecting the impact of foreground area on scattering intensity), the overexposure amplitude influence parameter (correcting the amplitude deviation caused by excessive exposure duration), and the phase shift parameter (derived from the optical path differences between foreground and scattering regions). These parameters are estimated once during the calibration phase and can be directly applied to subsequent depth correction. The proposed calibration process is straightforward, requires no hardware modification or complex modeling, and is applicable to a wide range of iToF scenarios. Experimental results across diverse scenes confirm that the method significantly improves depth accuracy in scattering-affected areas while maintaining high adaptability. Our overall calibration process is shown in Figure 2.

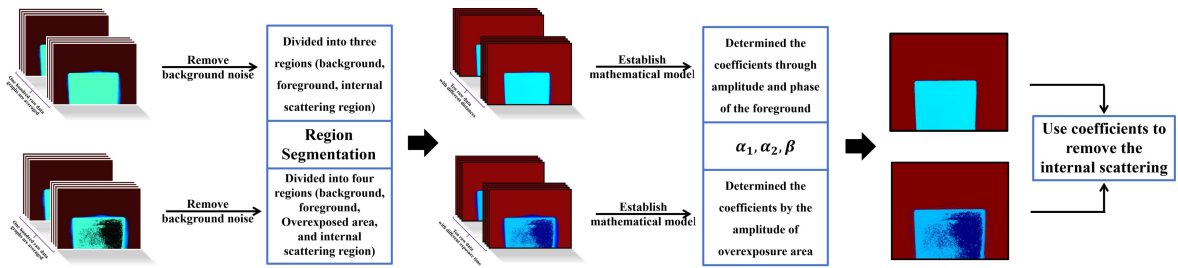

Fig. 2. Flowchart of the Overall Calibration Coefficients for Internal Scattering Interference Removal.

## 2. Principle

The indirect Time-of-Flight (iToF) camera emits modulated infrared light through the transmitter (TX) to uniformly illuminate the target object. The receiver (RX) captures the reflected light and analyzes its phase shift to calculate the time of flight of the light. The overall system is shown in Figure 3.

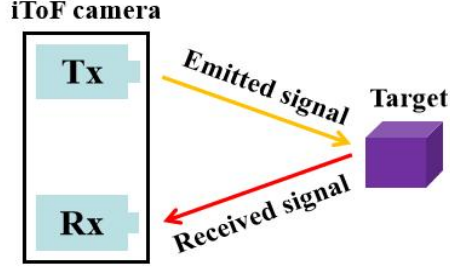

Fig. 3. Principle of indirect Time-of-Flight (iToF) camera.

The iToF camera estimates the distance of the target object based on the phase data to achieve depth sensing. The depth calculation usually employs the four-phase method [2], whose mathematical expression is as follows:

$$d = \frac{c}{4\pi f} \arctan\left(\frac{I_4 - I_2}{I_1 - I_3}\right) \quad (1)$$

where c is the speed of light in the vacuum. The f represents the modulation frequency. The I represents the light intensity received by the iToF camera under different modulation phases. Building on this, the values of the internal scattering region $I_{total}$ are influenced by the background signal $I_{bg}$ as well as internal scattering interference, which is generally represented by $I_{scatter}$. This relationship can be simplified to the following equation:

$$I_{total} = I_{bg} + I_{scatter} \quad (2)$$

where $I_{bg}$ is the ideal background pixel value, and $I_{scatter}$ represents the scatter generated by the foreground of interference. The light intensity composition of the internal scattering region, can be used to eliminate internal scattering interference. Based on the preceding light intensity decomposition, it is clear that accurately separating the foreground-induced scattering component $I_{scatter}$ is critical for restoring background depth.

Multiple sets of depth images are collected under controlled experiments by altering the position of the foreground object and adjusting exposure time to simulate different levels of internal scattering. The dataset covers both normally exposed and overexposed cases, indicating sensor saturation under current exposure conditions. We perform pixel-wise segmentation to extract the foreground, background, scattering, and overexposed regions. To meet real-time demands, we adopt a fast segmentation method based on adaptive thresholding and morphological operations.

During the modeling stage, we introduce a complex-valued light intensity formulation where the total received signal is expressed as a superposition of background and scattering components:

$$I_{total} = I_{bg} + I_{scatter} = A_{total}e^{i\phi_{total}} = A_{bg}e^{i\phi_{bg}} + A_{scatter}e^{i\phi_{scatter}} \quad (3)$$

Here, $A_{bg}, \phi_{bg}$ denote the amplitude and phase of the background signal, while $A_{scatter}$, $\phi_{scatter}$ denote those of the scattering signal. The modeling is decomposed into two components: phase shift modeling and amplitude perturbation modeling. The phase component is based on the optical path difference between the foreground and scattering paths. The foreground path is defined as:

$$L_{fg} = 2d_{fg} \quad (4)$$

while the scattering path is modeled as:

$$L_{scatter} = L_{fg} + \Delta L \quad (5)$$

According to the diffusion theory of the scattering optical path, the optical path difference caused by multiple scattering events can be approximated by a geometric function of the diffusion region [24], the optical path increment $\Delta L$ satisfies:

$$\Delta L \propto \sqrt{S_{scatter} + S_{fg}} - \sqrt{S_{fg}} \quad (6)$$

Here, $S_{fg}$, $S_{scatter}$ represent the areas of the foreground region and the internal scattering region, respectively. Assuming linear phase dependence on optical path, we introduce a scattering phase shift coefficient β, yielding:

$$\phi_{scatter} = \phi_{fg} + \frac{4\pi f}{c} \cdot \beta \cdot \left(\sqrt{S_{fg} + S_{sactter}} - \sqrt{S_{fg}}\right) \quad (7)$$

All area terms are obtained from the segmentation, while β is fitted using multi-sample experimental data.

In amplitude modeling, we assume the scattering energy originates from two sources: the normally exposed and overexposed regions. Overexposed areas are defined as pixels with grayscale value equal to 255, indicating saturation. The total scattering amplitude is decomposed as:

$$I_{scatter} = I_{normal} + I_{over\exp osure} \quad (8)$$

Two gain coefficients $\alpha_1$ and $\alpha_2$ are introduced, and a unified amplitude model is established after area normalization:

$$A_{scatter} \cos\phi_{scatter} = \overline{A_{normal}} \cdot \alpha_1 \cdot \frac{S_{scatter}}{S_{normal}} \cos\phi_{scatter} + \overline{A_{overexposure}} \cdot \alpha_2 \cdot \frac{S_{scatter}}{S_{overexposure}} \cos\phi_{scatter} \qquad (9)$$

Here, $S_{normal}$, $S_{overexposure}$ are the areas of the normally exposed and overexposed regions. With the three interpretable parameters $\alpha_1, \alpha_2, \beta$, the scattering modeling problem is formulated as a joint optimization task. These parameters can be estimated once during the calibration stage using limited data and then directly applied to correct depth errors in subsequent measurements. The final corrected background depth is expressed as:

$$d_{bg} = \frac{c}{4\pi f} \arctan\left( \frac{A_{total} \sin\phi_{total} - \left( \overline{A_{normal}} \cdot \alpha_1 \cdot \frac{S_{scatter}}{S_{normal}} + \overline{A_{overexposure}} \cdot \alpha_2 \cdot \frac{S_{scatter}}{S_{overexposure}} \right) \sin\left( \phi_{fg} + \frac{4\pi f}{c} \cdot \beta \cdot \left( \sqrt{S_{fg} + S_{scatter}} - \sqrt{S_{fg}} \right) \right)}{A_{total} \cos\phi_{total} - \left( \overline{A_{normal}} \cdot \alpha_1 \cdot \frac{S_{scatter}}{S_{normal}} + \overline{A_{overexposure}} \cdot \alpha_2 \cdot \frac{S_{scatter}}{S_{overexposure}} \right) \cos\left( \phi_{fg} + \frac{4\pi f}{c} \cdot \beta \cdot \left( \sqrt{S_{fg} + S_{scatter}} - \sqrt{S_{fg}} \right) \right)} \right) \qquad (10)$$

Based on the above formula, the calibration parameters $\alpha_1, \alpha_2, \beta$ can be directly used for real-time correction of depth images captured by iToF cameras. Specifically, the iToF camera first captures the raw depth map and amplitude map, both containing scattering effects. From the amplitude map, the normally exposed regions and overexposed regions can be identified, and the corresponding amplitude influence factors are calculated using the calibrated parameters. Next, the pre-calibrated phase shift parameter is used to compensate for the phase offset caused by scattering, reducing the impact of optical path differences on depth estimation. Finally, these amplitude and phase shift values are substituted into the correction formula to obtain accurate background depth, effectively suppressing internal scattering effects. Since these calibration parameters can be directly applied after calibration without requiring re-estimation during each measurement, this approach significantly reduces computational overhead and ensures high measurement accuracy under various complex lighting conditions.

## 3. Clibration Experiment

In the experiment, a Sony IMX570 CMOS dual-tap sensor, capable of measuring distances up to 10 meters, was used for data acquisition. To minimize background noise and ensure accurate parameter estimation, a low-reflectivity black curtain was selected as the background, while a high-reflectivity white cube was used as the foreground for calibration. This setup helps create a high-contrast environment, effectively isolating the scattering effects and reducing interference from unwanted reflections. The experimental system, including the iToF camera, is illustrated in Figure 4. The iToF camera was mounted on a movable platform, allowing precise adjustment of the relative position between the camera and the foreground object. The background-to-camera distance was fixed at 1 meter, while the initial foreground distance was set to 200 mm. The initial exposure time was set to 200 microseconds, providing a baseline for the normal exposure condition.

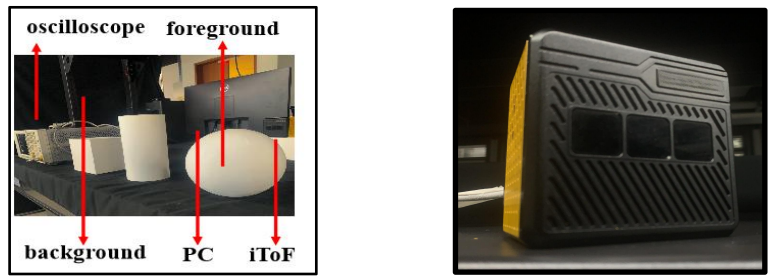

Fig. 4. The image on the left displays the iToF camera, developed by our research group, while the right Figure presents the overall system diagram of the experiment.

During each experimental round, the foreground object was moved forward in 50 mm increments, simulating different object distances, while the exposure time was increased by 10 microseconds per step to introduce controlled overexposure levels. At each position, the corresponding depth image was recorded. To reduce the impact of random noise, 100 repeated measurements were taken under each condition, and the average depth values were calculated to ensure accurate parameter fitting. The reduction of random noise is shown in Figure 5.

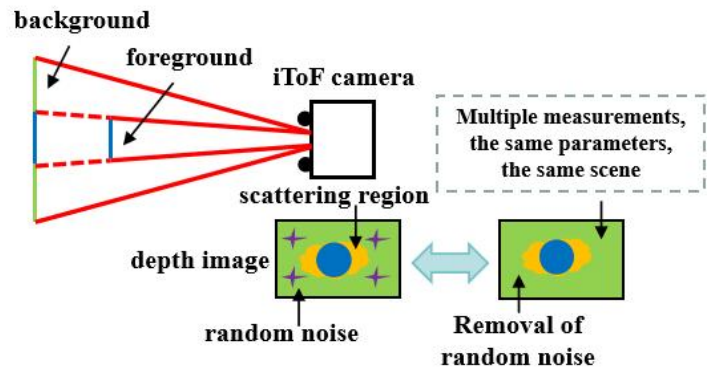

Fig. 5. Principle of Background Random Noise Removal

In the experimental image processing, a segmentation method combining grayscale thresholding and spatial distribution features was adopted to accurately extract the foreground (white cube), background (black curtain), and overexposed regions. This method includes image preprocessing, grayscale thresholding, spatial distribution feature analysis, and post-processing, ensuring accurate identification and separation of different regions. Additionally, depending on the performance and computational power of the iToF camera, more advanced algorithms such as k-means clustering [25] or Lightweight Neural Network [26] can be introduced to further improve segmentation accuracy. The extraction results are shown in the Figure 6 below.

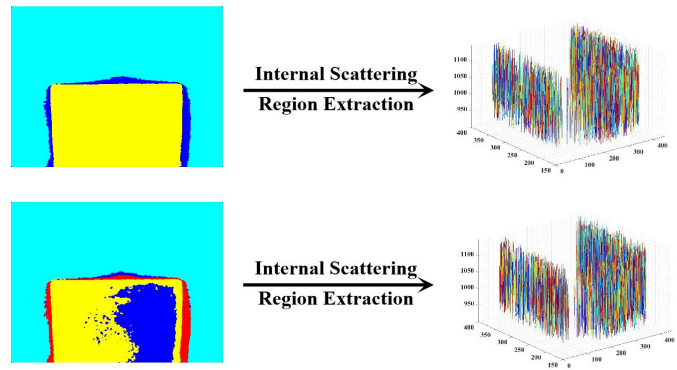

Fig. 6. The upper row represents the normal exposure condition, while the lower row shows the overexposure condition. The left side presents the segmentation results, and the right side displays the 3D visualization of the extracted internal scattering region.

The experimental process was divided into two distinct stages: segmented calibration under normal exposure and overexposure conditions. This approach was designed to separately capture the influence of normal exposure and overexposure on internal scattering, allowing for more precise parameter estimation [27-28]. The rationale behind this division lies in the underlying physical principle: exposure time essentially determines the amount of light reaching the CMOS sensor, i.e., the accumulated intensity over a given time period. Under fixed conditions such as object distance and illumination intensity, increasing the exposure time is equivalent to increasing the total energy received by the system. Therefore, by adjusting the exposure time, we are effectively controlling the strength of the internal scattering component during image formation. Normal exposure is used to model scattering behavior under typical imaging conditions, while overexposure amplifies the scattering effect, enabling more stable and comprehensive parameter mapping. In the normal exposure stage, the exposure time was fixed at 200 microseconds to prevent saturation in the foreground region, ensuring that the collected data accurately reflected the scattering effects without introducing signal clipping. During this phase, the foreground object was gradually moved from 200 mm to 300 mm in increments of 10 mm, covering a wide range of typical foreground distances. This stepwise approach enabled the collection of depth data at varying distances, providing a robust dataset for parameter fitting. Based on the statistical analysis of all captured images, the least squares method was then applied to fit the parameters, resulting in stable estimates of the amplitude influence coefficient $\alpha_1$ (corresponding to the left panel of Fig. 6) and the scattering phase shift coefficient $\beta$ (middle panel of Fig. 6) during the normal-exposure stage. In the subsequent overexposure stage, the foreground position was fixed at 200 mm to maintain a consistent reference point, while the initial exposure time was set to 250 microseconds. The exposure time was then incrementally increased by 10 microseconds per step until the foreground region became fully overexposed, effectively capturing the progressive impact of overexposure on scattering behavior. This approach allowed for a systematic investigation of the relationship between exposure duration and scattering amplitude. Based on this data, the amplitude influence coefficient $\alpha_2$ under overexposure conditions was then estimated (as shown in the right panel of Fig. 6), capturing the non-linear effects of extreme brightness on the internal scattering process. The results of this two-stage calibration process, including the fitted parameters, are presented in Figure 7.

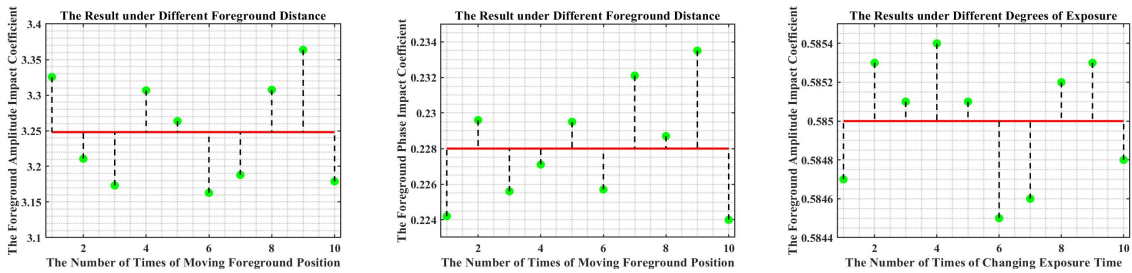

Fig. 7. The amplitude influence coefficient under normal exposure, the amplitude influence coefficient under overexposure, and the scattering phase shift coefficient obtained through least squares fitting.

To further validate the practical effectiveness of the proposed model, a series of experiments were conducted on the calibrated coefficients under multiple foreground distances (210 mm to 250 mm) and different exposure settings (200 to 290 microseconds), as shown in Figure 8. These tests were designed to assess the robustness and adaptability of the model under various real-world conditions, including different object placements and lighting intensities. The left plot in Figure 8 illustrates the prediction error rates after applying the calibration model proposed in this paper. It can be observed that all curves maintain an error rate below 10% in most cases, indicating that the model consistently provides accurate depth correction across a wide range of configurations. This demonstrates the model's strong stability and generalization capability, effectively compensating for scattering-induced errors. In contrast, the right plot presents the original, uncorrected scattering errors, which are significantly higher than the corrected results. This comparison further confirms that the proposed method effectively reduces scattering errors and significantly improves the ranging accuracy of iToF cameras under complex lighting conditions.

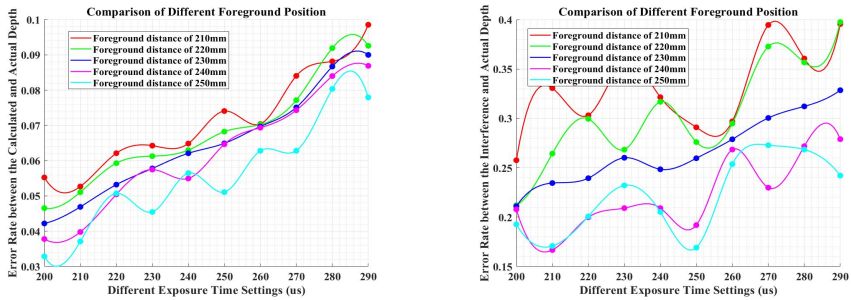

Fig. 8. The performance of the depth error between the calibrated internal scattering region and the ground truth, as well as that of the uncalibrated internal scattering region and the ground truth, under varying foreground distances and exposure times.

## 4. Result and Comparison

The proposed calibration model was applied to six representative real-world scenarios to further evaluate its robustness and generalization capability. These scenarios were carefully designed to cover a variety of configurations, including different numbers of foreground objects, varying background distances, and diverse surface textures, thereby reflecting the imaging diversity commonly encountered in practical iToF applications. Specifically, the test cases include: (1) a single foreground object with a nearby background, (2) a single foreground object with a distant background, (3) multiple foreground objects with a nearby background, (4) multiple foreground objects with a distant background, (5) a human palm, and (6) the back of a human hand. Each scenario was tested under both normal and overexposure conditions to assess the model's adaptability to different illumination intensities.

The visualization results are presented in Figure 9-14, where each scenario is displayed over two consecutive rows: the odd-numbered rows correspond to normal exposure, while the even-numbered rows correspond to overexposure. From left to right, each column shows the amplitude map, the raw depth map captured by the iToF camera, and the refined depth map obtained after applying the proposed internal scattering removal algorithm. This structured layout facilitates a clear comparison of the model's performance under different exposure levels and processing stages. The results demonstrate that, even under overexposure or complex scene compositions, the proposed method effectively suppresses scattering-induced depth errors and restores spatially coherent depth information, validating its applicability and robustness across a wide range of real-world environments.

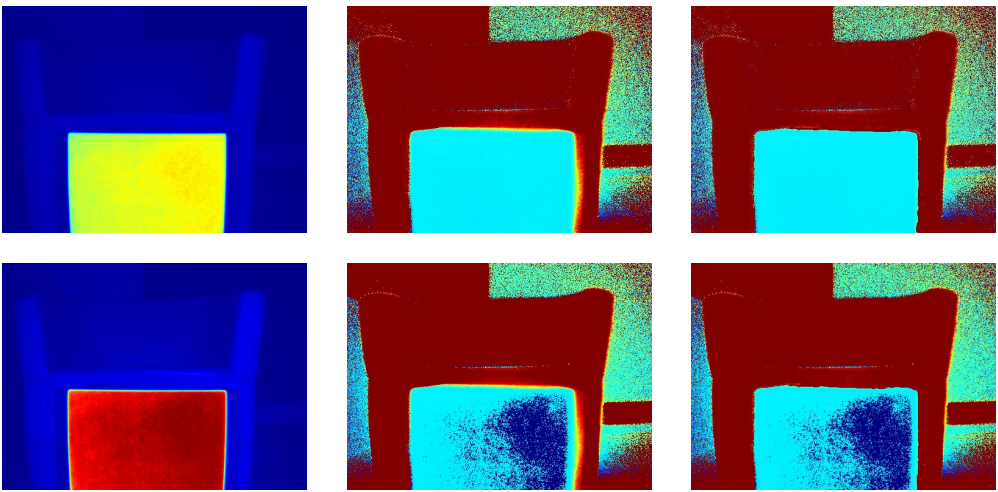

Fig .9. The first group shows the visual results of a single foreground against a relatively near background under normal exposure (odd rows) and overexposure (even rows) conditions.

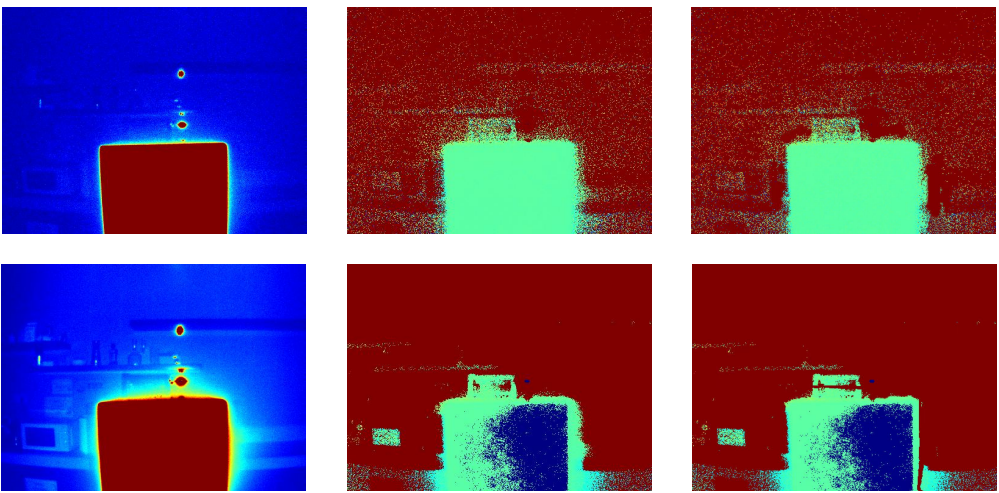

Fig .10. The second group shows the visual results of a single foreground against a relatively far background under normal exposure (first row) and overexposure (second row) conditions.

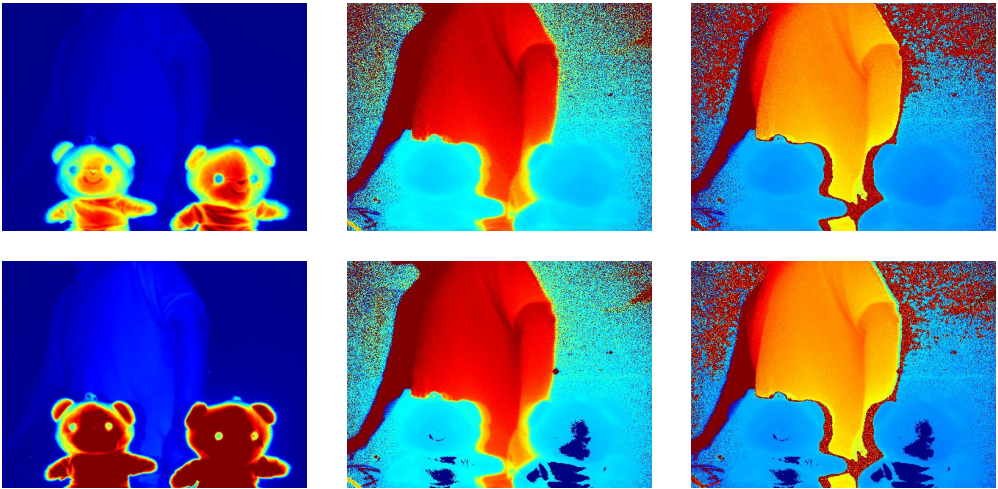

Fig .11. The third group shows the visual results of multiple foregrounds against a relatively near background under normal exposure (first row) and overexposure (second row) conditions.

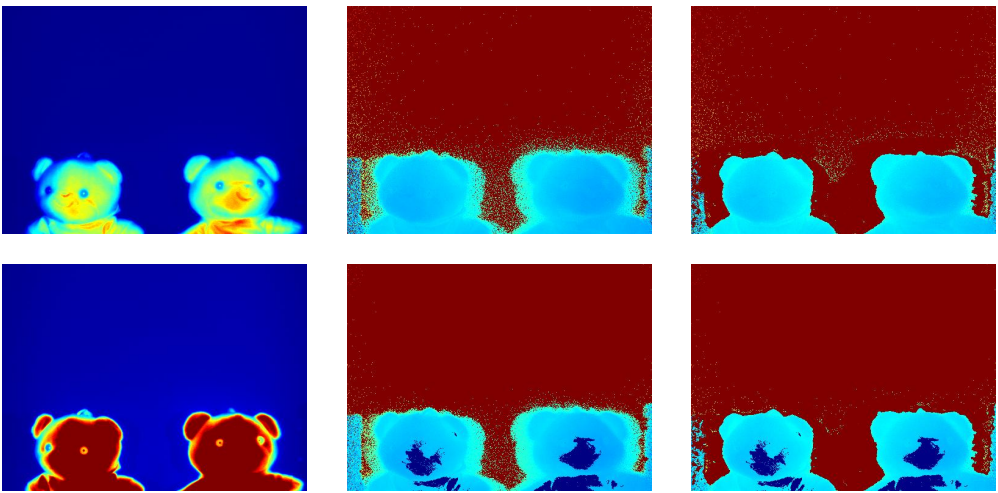

Fig .12. The fourth group shows the visual results of multiple foregrounds against a relatively far background under normal exposure (first row) and overexposure (second row) conditions.

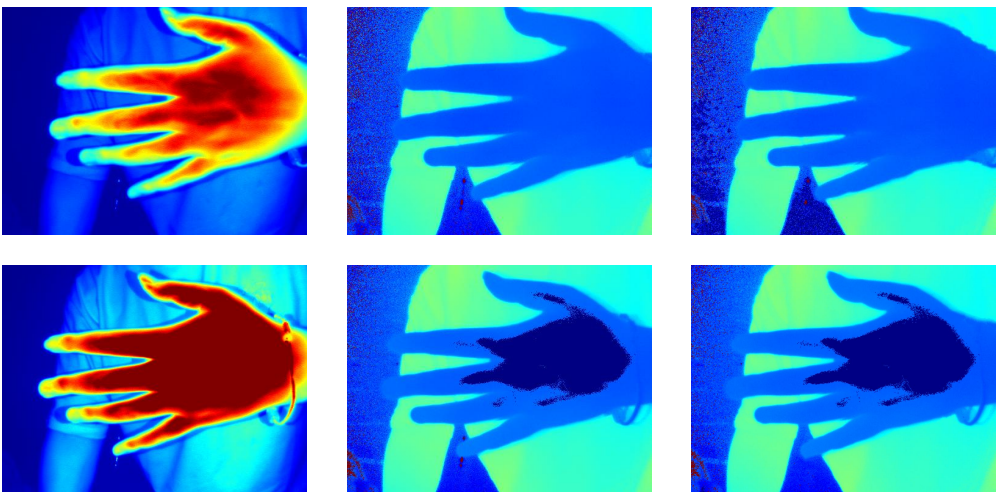

Fig .13. The fifth group shows the visual results of of a human palm under normal exposure (first row) and overexposure (second row) conditions.

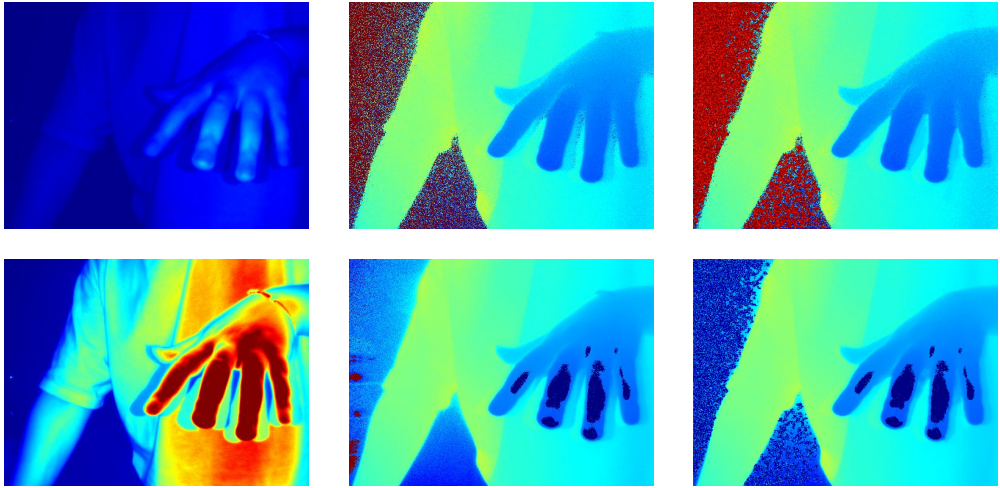

Fig .14. The sixth group shows the visual results of of a human back under normal exposure (first row) and overexposure (second row) conditions.

After calculation and analysis, the internal scattering interference in the above scenarios has been significantly suppressed, with the corrected error rate notably reduced. Figure 15 presents a comparison of depth error rates before and after calibration across six representative scenarios, where the left subplot shows the evaluation results under normal exposure, and the right subplot shows the results und er overexposure. Groups 1 to 6 correspond to the six test scenarios illustrated in Figure 8. Under normal exposure, the average depth error rate across the six groups decreased from 39% to 10.4% after correction; under overexposure, the average error rate was reduced from 47.5% to 12.4%.

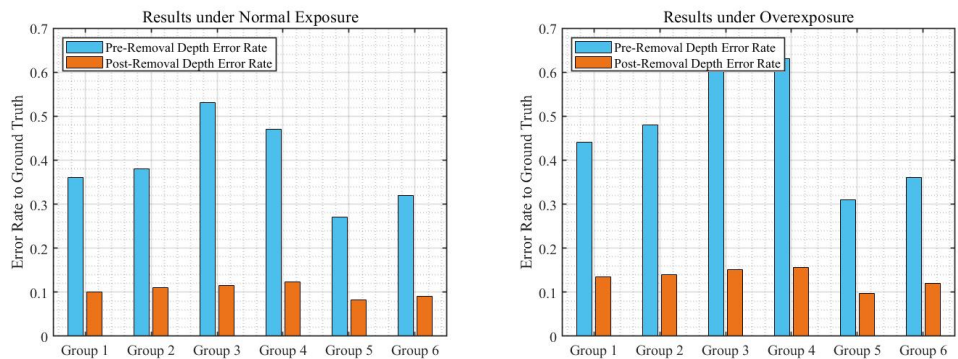

Fig. 15. Depth error rates before and after correction across six scenarios under normal (left) and overexposure (right) conditions. Groups 1-6 correspond to the scenes in Figure 8. Blue bars show pre-correction errors; orange bars show post-correction results.

To further validate the effectiveness of our proposed method, we also compare it against the Point Spread Function (PSF) model [9], which is widely adopted in the calibration of iToF systems. The PSF approach is commonly used due to its ability to simulate optical scattering and blurring processes. As shown in Figure 10, we conduct a comprehensive evaluation under both normal and overexposed conditions using three commonly used metrics: Mean Absolute Error (MAE), Root Mean Square Error (RMSE), and Structural Similarity Index (SSIM). The mathematical definitions are as follows:

·**Mean Absolute Error (MAE):**

$$MAE = \frac{1}{N}\sum_{i=1}^{N}\left|\hat{d}_i - d_i\right| \qquad (11)$$

where $\hat{d}_i, d_i$ denote the predicted and ground truth depth values, and N is the number of valid pixels.

·**Root Mean Square Error (RMSE):**

$$RMSE = \sqrt{\frac{1}{N}\sum_{i=1}^{N}\left(\hat{d}_i - d_i\right)^2} \qquad (12)$$

·**Structural Similarity Index (SSIM):**

$$SSIM = \frac{2\mu_{\hat{d}}\mu_d + c_1}{\mu_{\hat{d}}^2 + \mu_d^2 + c_1} \cdot \frac{2\sigma_{\hat{d}d} + c_2}{\sigma_{\hat{d}}^2 + \sigma_d^2 + c_2} \qquad (13)$$

where $\mu_{\hat{d}}$ and $\mu_d$ are the local means, $\sigma_{\hat{d}}^2$ and $\sigma_d^2$ are the local variances, $\sigma_{\hat{d}d}$ is the local covariance between the reconstructed depth $\hat{d}$ and the ground truth $d$ and $c_1, c_2$ are constants used to avoid instability when the denominators are small.

As illustrated in Figure 16, under normal exposure conditions, our method reduces the average MAE by 56.4% (from 13.6 mm for PSF to 5.93 mm), and the RMSE by 61.5% (from 40.3 mm to 15.5 mm). Meanwhile, the SSIM is improved by 36.2% (from 0.69 to 0.94), indicating a significant enhancement in structural preservation. Under overexposed conditions, the advantages of our method become even more pronounced: the average MAE is reduced by 66.3% (from 27.9 mm to 9.4 mm), RMSE drops by 65.6% (from 73.1 mm to 25.1 mm), and SSIM increases by 72.3% (from 0.49 to 0.84).

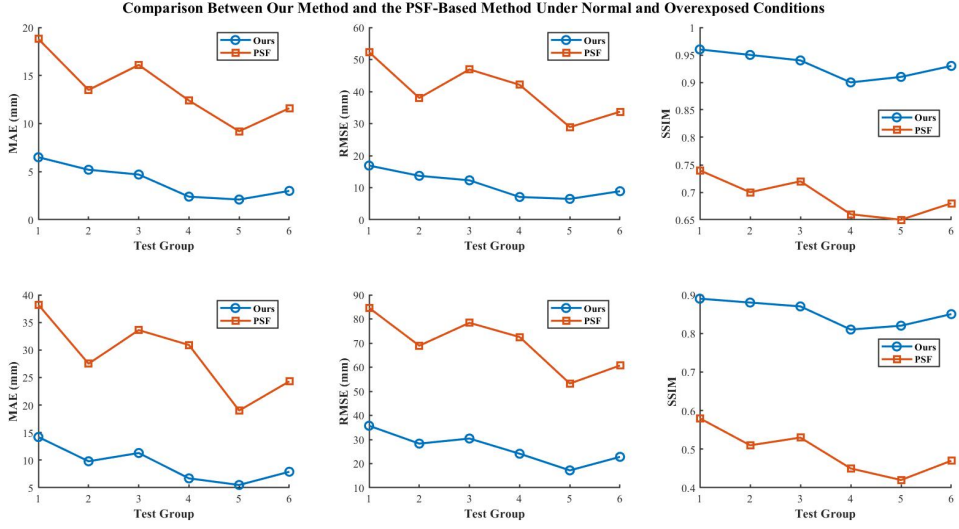

Fig. 16. Comprehensive evaluation of MAE, RMSE, and SSIM across six groups comparing our method and PSF calibration under varying exposure levels.

The experimental results above demonstrate that our method not only achieves depth reconstruction with higher fidelity to ground truth, but also significantly outperforms the traditional PSF model in terms of error suppression and structural consistency, exhibiting superior overall performance. These advantages enable our approach to maintain stable and reliable depth estimation under complex interference and varying exposure conditions, making it highly practical for real-world iToF system calibration and reconstruction tasks.

## 5. Conclusion

In this study, we propose a physically interpretable calibration method aimed at reducing depth errors caused by internal scattering in iToF cameras, particularly for depth correction under both normal exposure and overexposure conditions. By modeling amplitude perturbations and phase shifts induced by optical path variations, we construct a compact calibration process that requires only a small number of representative scenes. Experimental validation demonstrates significant improvements in depth measurement accuracy under various settings, confirming the method's strong generalization and robustness.

However, the proposed method still faces challenges when overexposure occurs in the background region or when all pixels of the foreground object are completely saturated. In such cases, the scattering region cannot be reliably segmented, making it difficult to estimate accurate amplitude-phase parameters and to perform quantitative correction. This limitation arises because the effective internal scattering zone is no longer distinguishable from saturated background signals, and the light intensity statistics become unreliable. These constraints hinder the applicability of our model in extreme scenes and limit its ability to generalize to highly saturated conditions.

Future work will focus on improving segmentation accuracy in complex environments, leveraging temporal information to handle dynamic lighting conditions, and exploring data-driven approaches to optimize parameter estimation under extreme exposure. We expect that combining these improvements with the current physical modeling approach will further enhance the model's generalization capability and real-time correction performance.

## Disclosures

There are no financial conflicts of interest to disclose.

## Code and Data Availability

The code associated with this paper is licensed and not publicly available. For specific questions regarding the code or access, please contact the authors directly. The paper's data are also processed using proprietary code; please reach out to the authors for any data-related inquiries.

## Acknowledgments

National Natural Science Foundation of China (U23A20282); Shenzhen Science, Technology and Innovation Commission (KJZD20240903095600001, JCYJ20220530142809022, WDZC20220811170401001).

## References

[1] Yansong Du, Zhaoxiang Jiang, Jindong Tian, and Xun Guan, "Modeling, analysis, and optimization of random error in indirect time-of-flight camera," Opt. Express 33, 1983-1994 (2025).

[2] S. Foix, G. Alenya, and C. Torras,"Lock-in Time-of-Flight (ToF) Cameras: A Survey," IEEE Sensors J. 11(9), 1917-926 (2011).

[3] X.Suand Q. Zhang,"Dynamic 3-D shape measurement method: a review,"Opt Lasers Eng 48(2), 191-204 (2010).

[4] A. Bhandari, C. Barsi, and R. Raskar, "Blind and reference-free fluorescence lifetime estimation via consumer time-of-flight sensors," Optica 2(11), 965 (2015).

[5] G.Wild, L.Fang, T. Newnham, A. Fisher, J. Palmer and C. Nagahawatt, "Comparative Performance of Simultaneous Localization and Mapping Algorithus for Unmanned Aircraft Based Navigation Systeus," 2018 5th IEEE International Workshop on Metrology for AeroSpace (MetroAeroSpace), Rome, Italy, 2018, pp. 123-127.

[6] James Mure-Dubois, Heinz Hügli, "Optimized scattering compensation for time-of-flight camera," Proc. SPIE 6762, Two- and Three-Dimensional Methods for Inspection and Metrology V, 67620H (10 October 2007).

[7] Lee, Sang Hun, et al. "850 nm IR transmissive electro-absorption modulator using GaAs micromachining." Sensors and Actuators A: Physical 197 (2013): 47-52.

[8] Zhou Y, Jiao F, Zeng P, et al. Inverse Designed SU8-on-Insulator Platform for Polymer Waveguide Photonics[J]. IEEE Photonics Technology Letters, 2025.

[9] Lichti, Derek D. et al. "New Models for Scattering Bias Compensation in Time-of-Flight Range Camera Self-Calibration." Journal of Surveying Engineering-asce 140 (2014): 04014003.

[10] Kavli, Tom et al."Modelling and compensating measurement errors caused by scattering in time-of-flight cameras."Optical Engineering + Applications (2008).

[11] Karel, Wilfried et al. "Modelling and Compensating Internal Light Scattering in Time of Flight Range Cameras." The Photogrammetric Record 27 (2012): n. pag.

[12] M. -S. Keel et al., "A VGA Indirect Time-of-Flight CMOS Image Sensor With 4-Tap 7-μm Global-Shutter Pixel and Fixed-Pattern Phase Noise Self-Compensation," in IEEE Journal of Solid-State Circuits, vol. 55, no. 4, pp. 889-897, April 2020.

[13] F. Gutierrez-Barragan, S. A. Reza, A. Velten and M. Gupta, "Practical Coding Function Design for Time-Of-Flight Imaging," 2019 IEEE/CVF Conference on Computer Vision and Pattern Recognition (CVPR), Long Beach, CA, USA, 2019, pp. 1566-1574.

[14] A. Islam, M. A. Hossain and Yeong Min Jang, "Interference mitigation technique for Time-of-Flight (ToF) camera," 2016 Eighth International Conference on Ubiquitous and Future Networks (ICUFN), Vienna,

Austria, 2016, pp. 134-139.

[15] Huijie Zhu and Hongwei Guo, "Joint least-squares algorithm correcting phase-shift errors and detector nonlinearity simultaneously in phase-shifting interferometry," Appl. Opt. 62, 1829-1839 (2023).

[16] D. Yang, D. An, Q. Wang, Z. Pan, and Y. Yue, "Scene Plane Recognition via Depth-Infrared Intensity Fusion Using ToF Camera," in CLEO 2024, Technical Digest Series (Optica Publishing Group, 2024), paper JTu2A.104.

[17] Talha Sultan, Syed Azer Reza, and Andreas Velten, "Towards a more accurate light transport model for non-line-of-sight imaging," Opt. Express 32, 7731-7761 (2024).

[18] Agresti, Gianluca and Pietro Zanuttigh. "Deep Learning for Multi-path Error Removal in ToF Sensors." ECCV Workshops (2018).

[19] H. Jung et al., "On the Importance of Accurate Geometry Data for Dense 3D Vision Tasks," 2023 IEEE/CVF Conference on Computer Vision and Pattern Recognition (CVPR), Vancouver, BC, Canada, 2023, pp. 780-791.

[20] M. Schelling, P. Hermosilla and T. Ropinski, "RADU: Ray-Aligned Depth Update Convolutions for ToF Data Denoising," 2022 IEEE/CVF Conference on Computer Vision and Pattern Recognition (CVPR), New Orleans, LA, USA, 2022, pp. 661-670.

[21] Y. Verdié, J. Song, B. Mas, B. Busam, A. Leonardis and S. McDonagh, "CroMo: Cross-Modal Learning for Monocular Depth Estimation," 2022 IEEE/CVF Conference on Computer Vision and Pattern Recognition (CVPR), New Orleans, LA, USA, 2022, pp. 3927-3937.

[22] D. Chen, T. Huang, Z. Song, S. Deng and T. Jia, "AGG-Net: Attention Guided Gated-convolutional Network for Depth Image Completion," 2023 IEEE/CVF International Conference on Computer Vision (ICCV), Paris, France, 2023, pp. 8819-8828.

[23] D. Qiu, J. Pang, W. Sun and C. Yang, "Deep End-to-End Alignment and Refinement for Time-of-Flight RGB-D Module," 2019 IEEE/CVF International Conference on Computer Vision (ICCV), Seoul, Korea (South), 2019, pp. 9993-10002.

[24] Najar, U., Barolle, V., Balondrade, P. et al. Harnessing forward multiple scattering for optical imaging deep inside an opaque medium. Nat Commun 15, 7349 (2024).

[25] T. Wahyuningrum, S. Khomsah, S. Suyanto, S. Meliana, P. E. Yunanto and W. F. Al Maki, "Improving Clustering Method Performance Using K-Means, Mini Batch K-Means, BIRCH and Spectral," 2021 4th

International Seminar on Research of Information Technology and Intelligent Systems (ISRITI), Yogyakarta, Indonesia, 2021, pp. 206-210.

[26] C. -H. Wang, K. -Y. Huang, Y. Yao, J. -C. Chen, H. -H. Shuai and W. -H. Cheng, "Lightweight Deep Learning: An Overview," in IEEE Consumer Electronics Magazine, vol. 13, no. 4, pp. 51-64, July 2024.

[27] Du Y, Deng Y, Zhou Y, et al. Multipath Interference Suppression in Indirect Time-of-Flight Imaging via a Novel Compressed Sensing Framework[J]. arXiv preprint arXiv:2507.19546, 2025.

[28] Shirvaikar M V. An optimal measure for camera focus and exposure[C]//Thirty-Sixth Southeastern Symposium on System Theory, 2004. Proceedings of the. IEEE, 2004: 472-47.

**Author biography**

Yansong Du is currently a Ph.D. candidate at Tsinghua University. His research focuses on novel camera design and computational imaging algorithms.

JingTong Yao is currently a Master candidate at Tsinghua University. Her research focuses on computer graphics as well as the modeling and algorithmic improvement of real-world physical problems.

Yuting Zhou is currently a Ph.D. candidate at Tsinghua University. His research interests include metasurfaces，photonic integration circuits, optical waveguides and sensors.

Feiyu Jiao is currently a Ph.D. candidate at Tsinghua University. His research focuses on performance analysis and algorithm design for visible light communication systems.

Dr. Zhaoxiang Jiang is an Associate Researcher at Guangdong Laboratory of Artificial Intelligence and Digital Economy (SZ), China. His research focuses on 3D sensing technologies.

Dr. Xun Guan is an Associate Professor at Tsinghua Shenzhen International Graduate School. His research interests in the field of optical communication and optical sensing, with a focus on wireless optical communication technology, optical fiber sensing technology, space optical perception, and integrated optical sensing technology.

**List of figure captions**

Figure 1: Schematic Diagram of Internal Scattering Interference Imaging.

Figure 2: Flowchart of the Overall Calibration Coefficients for Internal Scattering Interference Removal.

Figure 3: Principle of indirect Time-of-Flight (iToF) camera.

Figure 4: The image on the left displays the iToF camera, developed by our research group, while the right Figure presents the overall system diagram of the experiment.

Figure 5: Principle of Background Random Noise Removal.

Figure 6: The upper row represents the normal exposure condition, while the lower row shows the overexposure condition. The left side presents the segmentation results, and the right side displays the 3D visualization of the extracted internal scattering region.

Figure 7: The amplitude influence coefficient under normal exposure, the amplitude influence coefficient under overexposure, and the scattering phase shift coefficient obtained through least squares fitting.

Figure 8: The performance of the depth error between the calibrated internal scattering region and the ground truth, as well as that of the uncalibrated internal scattering region and the ground truth, under varying foreground distances and exposure times.

Figure 9: Visual results of six real-world groups under normal (odd rows) and overexposure (even rows) conditions. Each pair of consecutive rows represents one scene under two exposure settings. From left to right: amplitude image, raw iToF depth, and corrected depth using our method. The results highlight the method's effectiveness in reducing scattering-induced depth errors across diverse lighting and geometric conditions.

Figure 10: Depth error rates before and after correction across six scenarios under normal (left) and overexposure (right) conditions. Groups 1-6 correspond to the scenes in Figure 8. Blue bars show pre-correction errors; orange bars show post-correction results.

Figure 11: Comprehensive evaluation of MAE, RMSE, and SSIM across six groups comparing our method and PSF calibration under varying exposure levels.